\documentclass[twocolumn,superscriptaddress,floatfix,preprintnumbers]{revtex4-2}
\usepackage{amsfonts}
\usepackage{physics}
\usepackage{amsmath}
\usepackage{float}
\usepackage{multirow}
\usepackage{graphicx}
\usepackage{siunitx}
\usepackage{hyperref}
\hypersetup{
	colorlinks   = true, 
	urlcolor     = blue, 
	linkcolor    = blue, 
	citecolor   = blue 
}

\newcommand{\Jex}{J_{\text{ex}}}

\newcommand{\lhf}{$\mathrm{LiHoF_4}$}
\newcommand{\lhfx}{$\mathrm{LiHo_{x}Y_{1-x}F_4}$}
\newcommand{\Ho}{$\mathrm{Ho^{3+}}$}
\newcommand{\Y}{$\mathrm{Y^{3+}}$}
\newcommand{\up}{\ket{\uparrow}}
\newcommand{\down}{\ket{\downarrow}}

\begin{document}

\title{The Effect of Intrinsic Quantum Fluctuations on the Phase Diagram of Anisotropic Dipolar Magnets}

\author{Tomer Dollberg}
\author{Juan Carlos Andresen}
\author{Moshe Schechter}
\affiliation{%
 Department of Physics, Ben-Gurion University of the Negev, Beer Sheva 84105, Israel
}

\date{\today}

\begin{abstract}
The rare-earth material \lhf{} is believed to be an experimental realization of the celebrated (dipolar) Ising model, and upon the inclusion of a transverse field $B_x$, an archetypal quantum Ising model. Moreover, by substituting the magnetic Ho ions by non-magnetic Y ions, disorder can be introduced into the system giving rise to a dipolar disordered magnet and at high disorders to a spin-glass.
Indeed, this material has been scrutinized experimentally, numerically and theoretically over many decades with the aim of understanding various collective magnetic phenomena. One of the to-date open questions is the discrepancy between the experimental and theoretical $B_x -T$ phase diagram at low-fields and high temperatures.
Here we propose a mechanism, backed by numerical results, that highlights the importance of quantum fluctuations induced by the off-diagonal dipolar terms, in determining the critical temperature of anisotropic dipolar magnets in the presence and in the absence of a transverse field. We thus show that the description as a simple Ising system is insufficient to quantitatively describe the full phase diagram of \lhf{}, for the pure as well as for the dilute system.
 \end{abstract}

\maketitle

\textit{Introduction\label{sec:intro}.}---Anisotropic dipolar magnets, realized in both single molecule magnets and rare earth magnetic insulators, are at the forefront of quantum research. The large anisotropy barrier allows their use as nano-magnets, with possible applications in the operation of qubits and memory bits at reduced sizes \cite{Bertaina_rare-earth_2007, Zhong_nanophotonic_2017, Moreno-pineda_measuring_2021}.
In lattice form, anisotropic dipolar magnets typically have very small exchange interactions, allowing for efficient induction of quantum fluctuations by applied transverse fields. Thus, anisotropic dipolar magnets are perceived as experimental models for the transverse field Ising model (TFIM). These unique characteristics motivated intense study of quantum phenomena in these materials, including quantum phase transitions \cite{Bitko_Aeppli_1996_PhysRevLett.77.940, Ronnow_2005_389, Burzuri_magnetic_2011, Libersky_direct_2021}, quantum annealing \cite{Brooke_quantum_1999, Burzuri_magnetic_2011, Saubert_microscopics_2021}, domain wall dynamics \cite{Suzuki_propagation_2005, Silevitch_switchable_2010}, and high-Q nonlinear dynamics \cite{Silevitch_tuning_2019}.

One of the most studied anisotropic dipolar magnets is \lhf{} \cite{Brooke_quantum_1999, Saubert_microscopics_2021, Wu_classical_1991,
Ronnow_2005_389}.  Below the Curie temperature of $T_c=\SI{1.53}{\K}$ \lhf{}
orders ferromagnetically due to the dipolar interaction between \Ho{} ions
combined with its lattice structure \cite{Cooke_ferromagnetism_1975}. By the
inclusion of an external transverse field the transition temperature is
suppressed, until eventually it is converted to a quantum
phase transition at $B_x \approx \SI{4.9}{\tesla}$
\cite{Bitko_Aeppli_1996_PhysRevLett.77.940}.  Additionally, disorder can be
introduced by randomly substituting some of the magnetic \Ho{} ions with nonmagnetic \Y{} ions,
resulting in $\mathrm{LiHo_{x}Y_{1-x}F_4}$, which presents a rich phase
diagram---including a spin-glass phase at low concentrations ($x \lesssim
0.25-0.3$) \cite{Reich_dipolar_1990, Tam_spin-glass_2009,
Andresen_Schechter_2013_PhysRevLett.111.177202}---the result of the interplay of
interactions, disorder, and quantum fluctuations
\cite{Schechter_Stamp_2005_PhysRevLett_95_267208,
Schechter_Laflorencie_2006_PhysRevLett.97.137204,
Tabei_Gingras_2006_PhysRevLett.97.237203, silevitch_2007_ferromagnet_nature,
Schechter_2008_PhysRevB.77.020401,Gingras_2011_J._Phys._Conf._Ser._320_012001,
Andresen_existence_2014}

The $B_x-T$ phase diagram of \lhf{} is indeed in qualitative agreement with that
of the transverse-field Ising model, but a quantitatively correct description
has proven enduringly elusive, specifically at the high-temperature, low-field
regime, where thermal, rather than quantum, fluctuations are dominant
\cite{Tabei_Gingras_2008_PhysRevB.78.184408,Gingras_2011_J._Phys._Conf._Ser._320_012001,
Dunn_Hill_experimental_2012_PhysRevB.86.094428}.

The significance of off-diagonal terms of the dipolar interaction is well appreciated in the presence of disorder and a transverse field, as they break $\mathbb{Z}_2$ symmetry and transform spatial disorder to an effective random longitudinal
field, making this material one of the few magnetic realizations of the random field Ising model \cite{Schechter_Laflorencie_2006_PhysRevLett.97.137204,Tabei_Gingras_2006_PhysRevLett.97.237203,silevitch_2007_ferromagnet_nature,
Schechter_2008_PhysRevB.77.020401,Gingras_2011_J._Phys._Conf._Ser._320_012001}.
In this paper we establish the importance of offdiagonal dipolar (ODD) interaction terms to the quantitative description of the phase diagram of \lhfx{} at $0.4 < x \leq 1$, which may thus provide insights to open questions in the field.
We show that even for the pure system, and in the absence of a transverse field, the classical Ising model, which does not take these terms into account, provides an insufficient description of the system.
The reason being that ODD interactions give rise to quantum fluctuations which markedly affect the phase diagram.
These fluctuations are induced when ODD terms exert internal transverse fields that lower the energy of the \Ho{} ions on which they are exerted. We argue such fields are more prevalent in the paramagnetic (PM) phase than in the ferromagnetic (FM) phase, thus favoring the former.

\begin{figure*}[t]
	\centering
	\includegraphics[width=0.95\columnwidth]{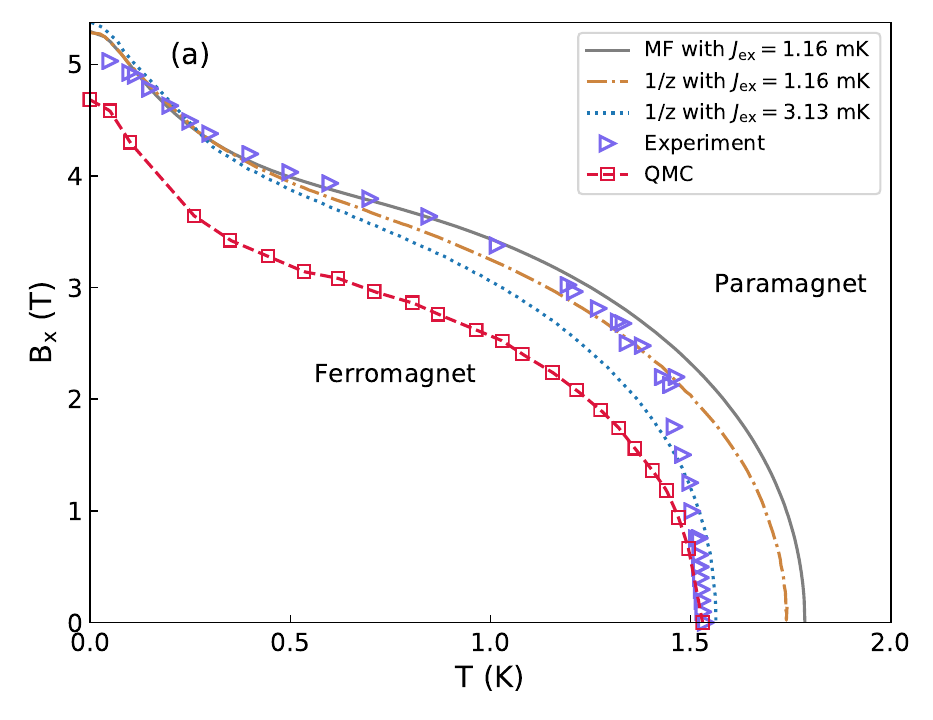}
	\quad
	\includegraphics[width=0.95\columnwidth]{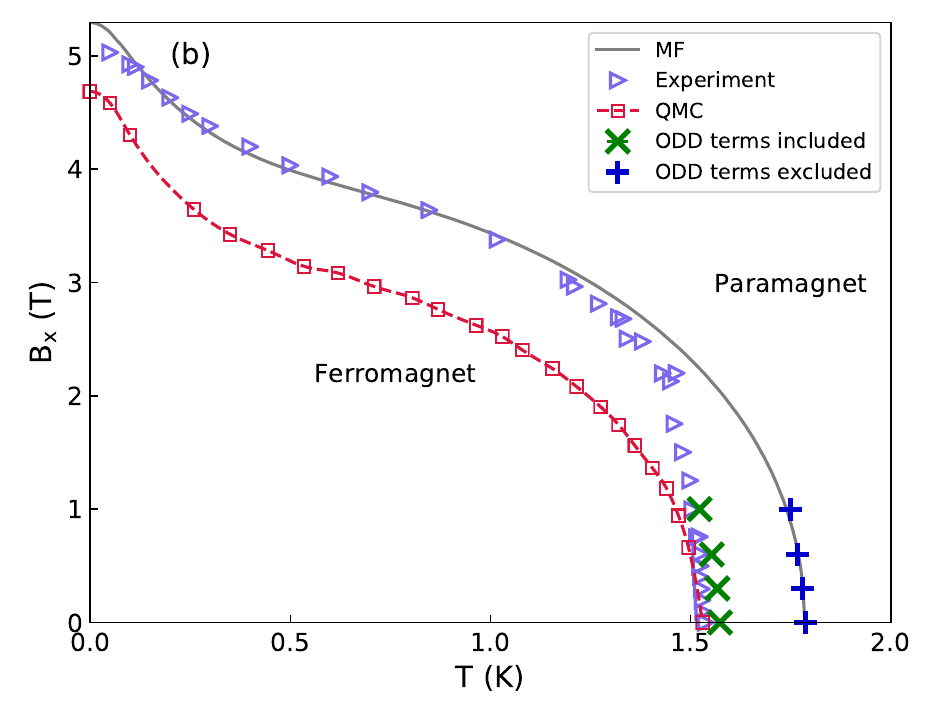}
	\caption{The full phase diagram of \lhf{} as a function of temperature and
		applied transverse field. (a) A compilation of previous numerical works. Open
		squares are quantum Monte Carlo (QMC) results
		\cite{Chakraborty_2004_PhysRevB.70.144411}. The dotted line is 1/z calculation
		with $\Jex=\SI{3.13}{\milli\kelvin}$ and the dot-dashed line is the same
		calculation with $\Jex=\SI{1.16}{\milli\kelvin}$
		\cite{Ronnow_2007_PhysRevB.75.054426}. The solid line is a mean-field
		calculation \cite{Babkevich_Ronnow_2016_PhysRevB.94.174443} which uses the
		latter exchange value. Triangles represent results from several different
		experiments
		\cite{Bitko_Aeppli_1996_PhysRevLett.77.940,
			Babkevich_Ronnow_2016_PhysRevB.94.174443, Dunn_Hill_experimental_2012_PhysRevB.86.094428}.
		An apparent trade-off is observed between theoretical predictions that match
		the experimental results at low temperatures but completely fail at the
		low-field regime, and ones that give correct zero-field $T_c$ but fail to
		predict the correct $T_c(B_x)$ dependence and give a poor match at the
		intermediate $B_x$ region. (b) Results of this paper overlaid on top of
		previous theoretical and experimental results.  The green Xs are the numerical results of this work with off-diagonal dipolar
		terms included and blue plus signs are for numerical results where they are excluded. Both use
		$\Jex=\SI{1.16}{\milli\kelvin}$.} \label{fig:phase-diagram}
\end{figure*}

Results from previous studies, using various Monte Carlo (MC) techniques
\cite{Chakraborty_2004_PhysRevB.70.144411,
	Tabei_Gingras_2008_PhysRevB.78.184408} and mean-field analyses
\cite{Bitko_Aeppli_1996_PhysRevLett.77.940,Ronnow_2007_PhysRevB.75.054426}, show
a persistent discrepancy with experimental results for the $B_x-T$ phase diagram
\cite{Bitko_Aeppli_1996_PhysRevLett.77.940,Gingras_2011_J._Phys._Conf._Ser._320_012001,
	Dunn_Hill_experimental_2012_PhysRevB.86.094428,
	Babkevich_Ronnow_2016_PhysRevB.94.174443}.  Namely, when the theoretical results
are fitted to the experimental results, either the zero-field critical
temperature or the low-temperature--high-field regime can be made compatible
with experiment, but not both.  If the former is chosen, then even at small
fields the $T_c(B_x)$ dependence is not theoretically well reproduced, and if
the latter, then the critical temperatures at low and intermediate transverse
fields are significantly overestimated.  The fitting is done by tuning the
single free parameter representing the nearest-neighbor antiferromagnetic
interaction strength, where higher values correspond to lower critical
temperatures.  This apparent trade-off can be clearly seen in
Fig.~\ref{fig:phase-diagram}(a).

Employing classical Monte Carlo simulations with variable single-spin magnetic moments, we find that the inclusion of ODD terms in the effective Hamiltonian allows for fitting $T_c$ at zero field using the same exchange parameter that accurately fits the data at low temperatures and high transverse fields.  At the same time, we find better agreement with experimental results for the long unexplained weak dependence of $T_c$ on the transverse field at small fields, and the linear dependence of $T_c$ on \Ho{} concentration in the absence of a transverse field.

\begin{figure}
	\centering
	\includegraphics[width=0.95\columnwidth]{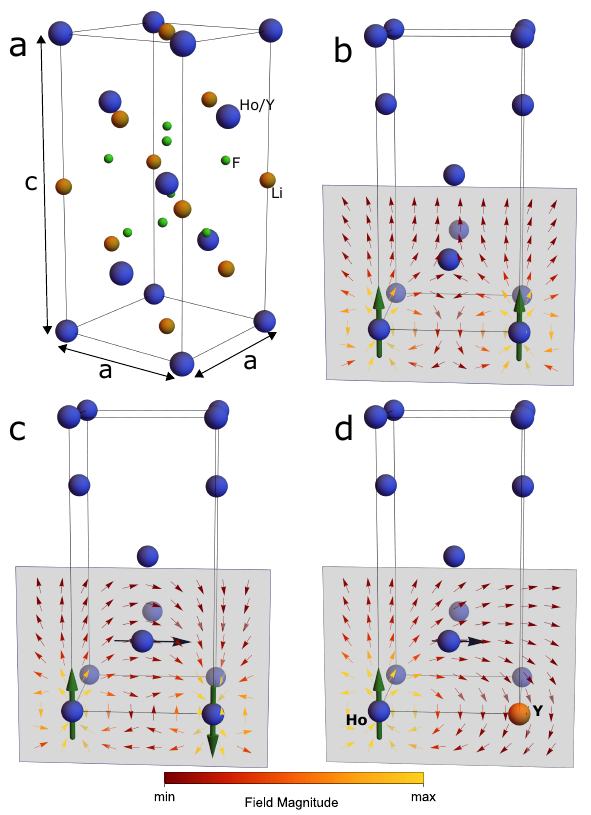}
  \caption[Schematic illustration of induced transverse local fields]{Crystal and magnetic structures of \lhfx{}.
  	(a) Crystal structure of \lhf{}. Only the $\mathrm{F^-}$ ions nearest to the central \Ho{} site are shown.
  	(b) Correlated spins, more abundant in the FM phase, induce a field with vanishing transverse component on their common intermediate neighbor.
  	(c) Spins in opposite orientations, more abundant in the PM phase, induce a nonzero transverse magnetic field on the third spin.
  	(d) When one of the \Ho{} ions is replaced by an \Y{} ion, the remaining Ho induces a transverse field on its neighbor regardless of its orientation.
	Thick green arrows indicate dipole moments, small colored arrows show the magnetic field generated by the two magnetic dipoles and a narrow black arrow qualitatively indicates the magnitude of the transverse component of the magnetic field exerted on the middle ion.
	The full effect captured in the simulation is a result of the transverse field not just on the middle ion, but on all other ions in the system. For further details on this point see Supplemental Material \cite{Note1}.}
	\label{fig:mechanism-sketch}
\end{figure}

\textit{\label{sec:Theory}Theoretical considerations.}---Off-diagonal terms of the dipolar interaction have been known to give rise to
many interesting phenomena in the case of the diluted
$\mathrm{LiHo_{x}Y_{1-x}F_4}$ in presence of an {\it external} transverse field, 
where they do not cancel by symmetry
\cite{Tabei_Gingras_2006_PhysRevLett.97.237203,
silevitch_2007_ferromagnet_nature, Schechter_Stamp_2008_PhysRevB.78.054438,
Schechter_2008_PhysRevB.77.020401, Schechter_Laflorencie_2006_PhysRevLett.97.137204}.
We argue that similar effects, arising from {\it internal} transverse fields exerted
by the single ion expectation values $\expval{J^z_i}$ on the $x$ angular momentum component
$J^x_j$ through terms of the form $V^{zx}_{ij} \expval{J^z_i}J^x_j$, make a
significant impact on the phase diagram even in the undiluted case.  The reason is that these ODD terms have a distinctly different contribution in the FM phase, where they are more likely to cancel by symmetry, than in the paramagnetic phase, where they are less likely to do so.
For example, a pair of spins that lie along the $a$-axis of the crystal will exert a transverse field on spins located between them, above or below the axis connecting the two spins, if the two spins have opposite orientations. Said field acts to lower the energy of the spin on which it acts regardless of its state, thereby energetically favoring the anti-aligned configuration of its two neighboring spins. See illustrations in Fig.~\ref{fig:mechanism-sketch}. This interaction thus constitutes a disorder-enhancing mechanism which acts to decrease the critical temperature. It requires the existence of three spins and correlation between two of them---an important aspect which will be discussed further below.

Another effect of the transverse fields is to decrease the absolute value of
$\expval{J^z}$ for the two lowest single-ion electronic energy states, by mixing
them with the higher electronic states. This also contributes to the reduction
of $T_c$ just by reducing the dominant $zz$ dipolar term proportional to
$\expval{J^z}^2$.  This mechanism of the correlation induced enhancement of the
transverse field, by its nature, is not likely to be captured
in any sort of mean-field-like analysis as it depends on the spatial
fluctuations of the states. We posit that including the ODD terms is necessary to
explain the previously mentioned discrepancies between theory and experiment. 

\textit{\label{sec:Numerical}Numerical details.}---In order to examine the effect of ODD terms on the phase diagram of \lhfx{} we perform Monte Carlo simulations using an effective Hamiltonian derived building upon the work of Chakraborty \textit{et al.} \cite{Chakraborty_2004_PhysRevB.70.144411}, but keeping the ODD terms. In this way we get an effective Hamiltonian,
\begin{equation}\label{eq:rearranged-hamiltonain-eff}
\begin{split}
H_{\text{eff}} &= \sum_i V_C(\hat{\boldsymbol{J}}_i) - g_L \mu_B \sum_i
  \hat{\boldsymbol{B}}_i\cdot\hat{\boldsymbol{J}}_i
\end{split}
\end{equation}
where
\begin{align}\label{eq:effective-B}
\hat{B}^x_i &= B_x - g_L \mu_B \sum_{j\neq i} V^{zx}_{ij} \hat{J}^z_j \nonumber\\
\hat{B}^y_i &= - g_L \mu_B \sum_{j\neq i} V^{zy}_{ij} \hat{J}^z_j\\
\hat{B}^z_i &= - \frac{1}{2} g_L \mu_B \sum_{j\neq i} V^{zz}_{ij} \hat{J}^z_j -
  \frac{\Jex}{2 g_L \mu_B} \sum_{j \in NN} \hat{J}^{z}_{j} \nonumber
\end{align}
act as effective internal fields when taking their expectation values, thereby
transforming Eq.~\eqref{eq:rearranged-hamiltonain-eff} to an effective Hamiltonian for the single spins $i$.
The $V_C(\boldsymbol{J}_i)$ term is a crystal field potential which imposes an Ising easy axis along the $c$ axis of the crystal, with a first excited state at $\sim \SI{10}{\kelvin}$ above the ground-state doublet
\cite{Ronnow_2007_PhysRevB.75.054426}.
$V_{ij}^{\mu \nu}$ is the magnetic dipole interaction, $\Jex$ is the nearest-neighbor
exchange interaction coupling constant, $\mu_B = \SI{0.6717}{\kelvin\per\tesla}$
is the Bohr magneton and $g_L=\frac{5}{4}$ is a Land\'{e} g-factor.
$\boldsymbol{J}_i$ are angular momentum operators of the \Ho{} ions.
See further details on the derivation of the effective Hamiltonian in the Supplemental Material \footnote{See Supplemental Material at [URL will be inserted by publisher] for further details on derivations and numerical methods and results.}.

Since the \Ho{} ions retain their Ising character up to transverse fields well
above the critical transverse field
\cite{Schechter_Stamp_2008_PhysRevB.78.054438} we model a single-ion as a
2-state Ising system under an applied field exerted by all other ions in the
system, as well as the external field.  This applied field not only shifts the
energies of the two states, but also modifies the single ion magnetic moments
$\expval{J^z}$ associated with them. These are the magnetic moments which in
turn exert magnetic fields on other ions.  Therefore, each site has two possible
states, $\up$ and $\down$, and each of these has two quantities of interest
associated with it: local energy $\bra{\uparrow(\downarrow)}
H_{\text{single-site}} \ket{\uparrow (\downarrow)}$ and magnetic moment
$\bra{\uparrow(\downarrow)} J^z \ket{\uparrow (\downarrow)}$ where 
\begin{equation}\label{eq:single-site-hamiltonian}
	H_{\text{single-site}} = V_C(\hat{\boldsymbol{J}}) - g_L \mu_B \boldsymbol{B} \cdot \hat{\boldsymbol{J}}.
\end{equation}
The method of employment of the effective Hamiltonian within the Monte Carlo simulations is detailed in the "Numerical Methods" section of the Supplemental Material \cite{Note1}.

\textit{\label{sec:Results}Results.}---Figure~\ref{fig:phase-diagram}(b) shows the $B_x-T$ phase diagram of \lhf{}. Our
results, with ODD terms included and excluded, both use the exchange parameter
$\Jex=\SI{1.16}{\milli\kelvin}$ suggested in Refs.
\cite{Ronnow_2007_PhysRevB.75.054426,Babkevich_Ronnow_2016_PhysRevB.94.174443} which corresponds to the fitting at low temperatures and high transverse fields.
It is easy to see that the simulation with ODD terms excluded corresponds to the
mean-field calculation, while the results with ODD terms included are in close
agreement with the experimental results, for zero and small transverse fields.

Thus, the inclusion of the ODD terms results in good agreement with experiment at zero transverse field without the need to choose $\Jex$ that is in clear disagreement with experiment at lower temperatures and higher transverse fields.
For finite but small transverse fields we find that the decrease in $T_c$ due to the ODD terms is maintained.  We emphasize that our simulation is a classical MC simulation, but one that allows for varying magnetic moments due to the influence of transverse fields.  Therefore, the $T_c(B_x)$ dependence in our results is a consequence of the renormalization of individual magnetic moments due to the quantum coupling of each of the Ising doublet states to higher excited electronic states, as opposed to quantum fluctuations between the two Ising doublet states.
Hence, our model is expected to be valid at low fields $B_x \lesssim \SI{1}{\tesla}$ where quantum fluctuations are small \cite{Schechter_Stamp_2008_PhysRevB.78.054438}.

\begin{figure}
	\centering
	\includegraphics[width=0.95\columnwidth]{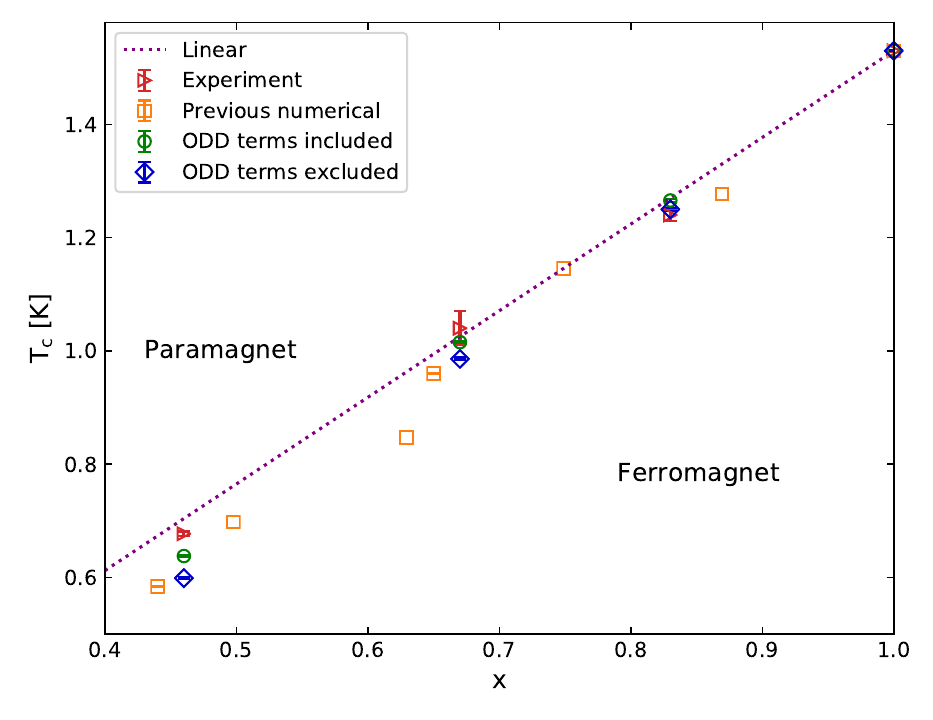}
	\caption[$x-T$ phase diagram of \lhfx{} with results]{Transition temperature,
		$T_c$, vs. \Ho{} concentration, $x$, phase diagram from different sources:
		experimental from Ref.~\cite{Babkevich_Ronnow_2016_PhysRevB.94.174443}
		(triangles), numerical from
		Refs.~\cite{Andresen_Schechter_2013_PhysRevLett.111.177202,Biltmo_Henelius_2007_PhysRevB.76.054423}
		(squares) and this work with ODD terms included (circles) and ODD terms
		excluded (diamonds).
		A dotted line shows the mean-field prediction (linear).
		Numerical results from this work are scaled so that they agree with the experimental $T_c(x=1)=\SI{1.53}{\kelvin}$. 
		The results with ODD terms included use $\Jex=\SI{1.16}{\milli\kelvin}$, while
		the results with ODD terms excluded use $\Jex=\SI{3.91}{\milli\kelvin}$, as
		suggested, e.g. in Ref.~\cite{Tabei_Gingras_2008_PhysRevB.78.184408}.}
	\label{fig:phase-results-x-T}
\end{figure}

Another facet of the incomplete quantitative understanding of this material has to do with the phase diagram of \lhfx{} as a function of the \Ho{} concentration $x$ at zero applied field. At moderate to high concentrations ($x \gtrsim 0.4$) experiments
show a linear dependence of $T_c$ on $x$, in agreement with the mean-field
prediction
\cite{Babkevich_Ronnow_2016_PhysRevB.94.174443,Reich_dipolar_1990,Quilliam_experimental_2012},
whereas the available numerical work seems to indicate a steeper decline of
$T_c$ as $x$ is reduced \cite{Biltmo_Henelius_2007_PhysRevB.76.054423, Andresen_Schechter_2013_PhysRevLett.111.177202}.

We note here that the inclusion of the ODD terms in the effective low energy
Hamiltonian of the system leads to an effective 3-spin interaction, proportional to the anti-correlation of two spins and the existence of the third. We thus expect this term to depend strongly on Ho concentration, allowing for its distinction from the excess antiferromagnetic exchange used to parameterize the system, and for better agreement with the experimental $x-T$ phase diagram. In Fig.~\ref{fig:phase-results-x-T} we present our results for $T_c$ as function of Ho concentration $x$ in the presence of ODD terms and exchange
parameter $\Jex=\SI{1.16}{\milli\kelvin}$, and in the absence of ODD terms and
$\Jex=\SI{3.91}{\milli\kelvin}$. Indeed, the results with ODD terms included show milder reduction of $T_c$ with decreased concentration, in better agreement with the experimental findings of $T_c(x)=x T_c(x=1)$.

\begin{figure}
	\centering
	\includegraphics[width=0.95\columnwidth]{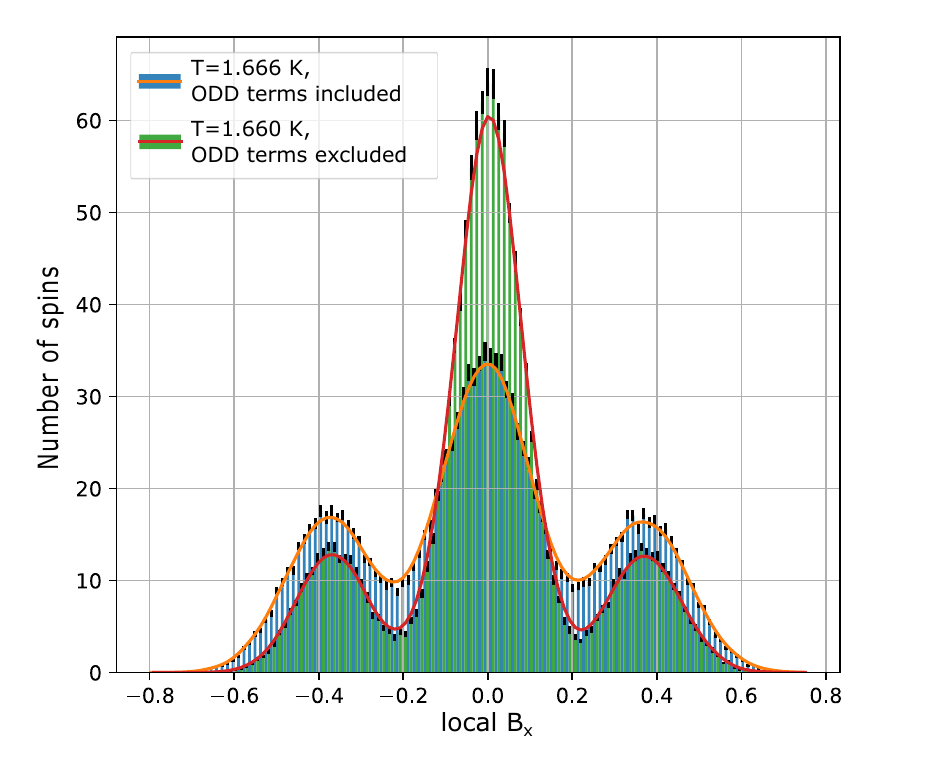}
  \caption{Distribution of local $B_x$ at thermodynamic equilibrium for system
  size $L=7$ with zero external transverse field and at $x=1$. The temperatures
  used, around $T=\SI{1.66}{\kelvin}$, are below $T_c$ when ODD terms are
  excluded and above $T_c$ when ODD terms are included. The black vertical lines
  at the end of each bar are standard errors. For each distribution a solid
  smooth line is plotted as a guide to the eye, obtained by convolution of the
  bin values with a Gaussian function.  Where ODD terms are excluded, they are
  nevertheless considered at the end of the simulation for obtaining these
  effective transverse fields.  One can clearly see that when ODD terms are
  included, the distribution of $B_x$ is wider than when they are excluded---the
  peak at $B_x=0$ is lower, compensated by higher values at the exterior.  This
  is an indication of the ODD induced mechanism at work. Configurations
  maximizing internal transverse fields become more energetically favorable and
  thus for any given temperature they are more common.}
	\label{fig:Bx-hist}
\end{figure}

An additional microscopic indication of the effect of ODD terms can be obtained
by inspecting the distribution of local transverse fields.
Figure~\ref{fig:Bx-hist} shows the distribution of $B_x$ at the end of the
simulation, when the system has reached thermodynamic equilibrium, for simulations where ODD terms are included and where they are
excluded, yet still considered at the end of the simulation for the calculation
of the effective transverse fields. It is clear that, at a given temperature,
when ODD terms are included the distribution of $B_x$ becomes wider.  This is
expected, since when ODD terms are included, configurations that maximize
internal transverse fields become more energetically favorable and are thus more
abundant at any given temperature.

\begin{figure}
	\centering
	\includegraphics[width=0.95\columnwidth]{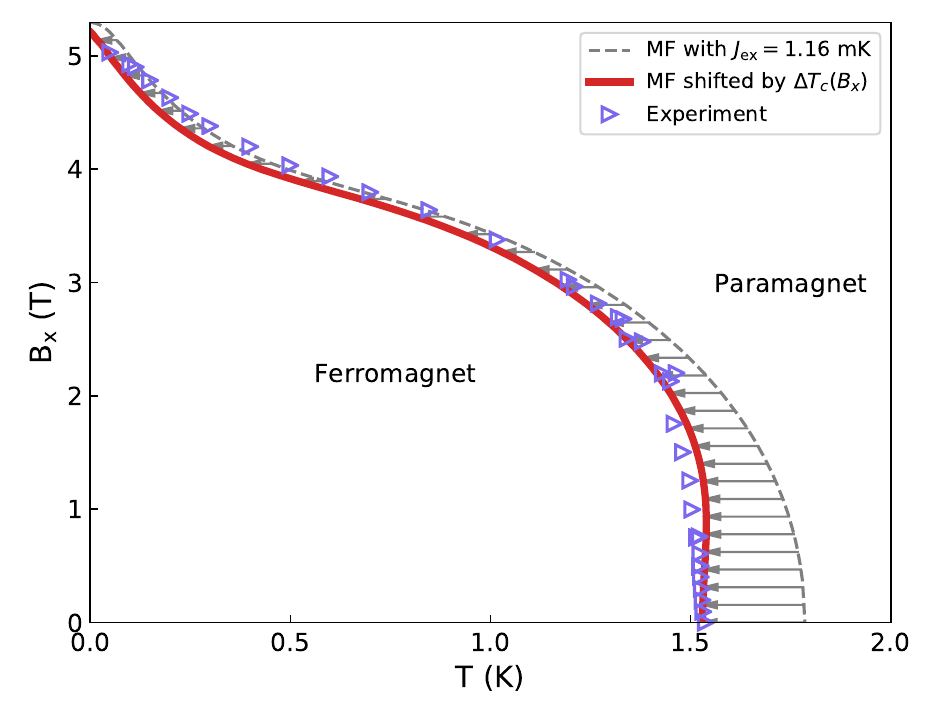}
	\caption{FM-PM phase transition line in the temperature-transverse field plane, as obtained by simple mean-field \cite{Babkevich_Ronnow_2016_PhysRevB.94.174443} (dashed gray line) and by including the $T_c$-reducing effect of ODD terms (solid red line).
		Horizontal gray arrows indicate the application of $\Delta T_c(B_x)$ to the mean-field result. Triangles denote the same experimental data also shown in Fig.~\ref{fig:phase-diagram}.}
	\label{fig:phase-diagram-shifted-Tc}
\end{figure}

Lastly, in order to demonstrate the effect of the ODD terms on the full $B_x-T$ phase diagram we pursue a simplified approach, assuming the main difference between the FM and PM phases relevant to the effectiveness of the ODD-induced mechanism is the width of the distribution of local transverse fields. For simplicity, we assume that the FM phase is characterized by a vanishing width of this distribution, while the PM phase is characterized by a finite width $h$ which we take to be $0.4\ \mathrm{T}$, i.e. the field exerted on a spin by its two nearest neighbors along the $x$ or $y$ directions while they are in opposite orientations from each other. Unsurprisingly, this is also approximately the value of the secondary peaks in Figure~\ref{fig:Bx-hist}. In both phases the distribution of local transverse fields is centered around the value of the external field $B_{x}$.
Hence, on average half of the spins experience a local transverse
field $B_{x}+h$ and the other half $B_{x}-h$. This is true even
though at $B_{x}>0$ the former option is more energetically favorable,
because, owing to the crystal's mirror symmetry, any spin which generates
a positive local field at some site also generates a negative local
field at another site.
Thus, we estimate the energy reduction of the PM phase due to the
ODD interactions, as $\Delta E(B_{x})=E(B_x)-\left[E(B_{x}-h)+E(B_{x}+h)\right]/2$,
where, for simplicity, we take $E(B_{x})$ to be the average of the
two lowest eigenenergies of \eqref{eq:single-site-hamiltonian} with the given $B_x$.
Assuming that the reduction in $T_{c}$ due to the inclusion of ODD
terms, $\Delta T_{c}$, is proportional to the energy reduction $\Delta E(B_{x})$, we find the appropriate factor by demanding $\Delta T_{c}(B_{x}=0)=T_{c}^{\mathrm{MF}}(0)-T_{c}^{\mathrm{exp}}(0)\approx1.79-1.53\approx0.25\ \mathrm{K}$.
Using this scaling factor we apply a $B_{x}$-dependent shift
$\Delta T_{c}(B_{x})$ to the mean-field phase boundary to obtain
an approximate phase boundary with the effect of ODD terms included,
as presented in Figure~\ref{fig:phase-diagram-shifted-Tc}.

\textit{\label{sec:Discussion}Discussion.}---We have shown here that the description of anisotropic dipolar
systems by the Ising model, and in the presence of a transverse field by the
transverse field Ising model, is essentially insufficient. Even for the pure
system off-diagonal dipolar terms induce an effective three spin interaction,
enhancing paramagnetic fluctuations and lowering the critical temperature. We
have analyzed the effect of the ODD terms on the relation between critical
temperature and both transverse field and dilution, thereby addressing unanswered puzzles regarding discrepancies between theory and experiment.  Our results at small fields are obtained with the same exchange parameter used to fit the phase transition at low temperatures and high transverse fields, and produce improved fitting to experimental data at finite transverse fields and as a function of Ho concentration. Thus, our results point to the need to include the quantum fluctuations induced by the off-diagonal terms in any theoretical consideration, classical and quantum, of anisotropic dipolar systems. Examples are classical and quantum annealing protocols, and a comprehensive quantum modeling of the system required to establish its full phase diagram.

\begin{acknowledgments}
We would like to thank Dror Orgad and Markus Müller for useful discussions. M.S. acknowledges support from the Israel Science Foundation (Grant No. 2300/19).

\end{acknowledgments}
%
%
%


%

\pagebreak
\widetext
\begin{center}
	\textbf{\large Supplemental Material: The Effect of Intrinsic Quantum Fluctuations on the Phase Diagram of Anisotropic Dipolar Magnets}
\end{center}
\setcounter{equation}{0}
\setcounter{figure}{0}
\setcounter{table}{0}
\setcounter{page}{1}
\makeatletter
\renewcommand{\thefigure}{S\arabic{figure}}
\renewcommand{\tablename}{Supplementary Table}
\renewcommand{\theequation}{S.\arabic{equation}}

\newcounter{SIfig}
\renewcommand{\theSIfig}{S\arabic{SIfig}}

\section{Effective Hamiltonian with offdiagonal dipolar (ODD) terms included}
Pure \lhf{} forms a tetragonal structure with lattice constants
$a=5.175\ \mbox{\normalfont\AA}$ and $c=10.75\ \mbox{\normalfont\AA}$, as shown in Fig. \ref{fig:mechanism-sketch}a of the main text. There are four \Ho{}
ions per unit cell which form a lattice with a basis with coordinates $(0, 0,
\frac{1}{2})$, $(0, \frac{1}{2}, \frac{3}{4})$, $(\frac{1}{2}, \frac{1}{2}, 0)$
and $(\frac{1}{2}, 0, \frac{1}{4})$
\cite{Gingras_2011_J._Phys._Conf._Ser._320_012001}.
The complete Hamiltonian of \lhf{} in a transverse magnetic field is given by
\cite{Chakraborty_2004_PhysRevB.70.144411,
	Gingras_2011_J._Phys._Conf._Ser._320_012001}
\begin{equation}\label{eq:micro-hamiltonian}
	H=\sum_i V_C(\boldsymbol{J}_i) - g_L \mu_B \sum_i B_x J_i^x + \frac{1}{2}
	(g_L \mu_B)^2 \sum_{i\neq j} V_{ij}^{\mu \nu} J_i^{\mu} J_j^{\nu} + \Jex
	\sum_{\left\langle i,j \right\rangle} \boldsymbol{J}_i \cdot
	\boldsymbol{J}_{j}
	+ A \sum_i (\boldsymbol{I}_i \cdot \boldsymbol{J}_i)
\end{equation}
where $V_{ij}^{\mu \nu}$ is the magnetic dipole interaction, $V_{ij}^{\mu \nu} =
\frac{\mu_0}{4\pi} \frac{\delta^{\mu \nu} |\vec{r}_{ij}|^2 - 3 (\vec{r}_{ij})^{\mu}
	(\vec{r}_{ij})^{\nu}}{|\vec{r}_{ij}|^5}$.  $\Jex$ is the nearest-neighbor
exchange interaction coupling constant. $\mu_B = \SI{0.6717}{\kelvin\per\tesla}$
is the Bohr magneton, $g_L=\frac{5}{4}$ is a Land\'{e} g-factor, and $\mu_0$ denotes vacuum permeability.
$\boldsymbol{J}_i$ are angular momentum operators of the \Ho{} ions. $A$ is the
hyperfine interaction strength, and $\boldsymbol{I}_i$ is the nuclear spin
operator, where the total $\mathrm{Ho}$ nuclear spin is $I=\frac{7}{2}$.  The
\Ho{} ions may be randomly substituted by nonmagnetic \Y{} ions to form \lhfx{}.
The crystal field term $V_C(\boldsymbol{J}_i)$ imposes an Ising easy axis along
the $c$ axis of the crystal, with a first excited state at $\sim
\SI{10}{\kelvin}$ above the ground-state doublet
\cite{Ronnow_2007_PhysRevB.75.054426}.

\label{sec:eff-hamiltonian-derivation}Following an approach analogous to that of Chakraborty \textit{et al.} \cite{Chakraborty_2004_PhysRevB.70.144411} we recast the full Hamiltonian \eqref{eq:micro-hamiltonian} as an effective transverse-field Ising model Hamiltonian, but keep the neglected ODD terms. Of the diagonal dipolar terms, we keep only the $zz$ interactions which have been established as the most dominant, but we also keep the off-diagonal 
interaction terms. The dipolar interaction is invariant under both $i \leftrightarrow j$ 
and $\mu \leftrightarrow \nu$. We also use the fact that $\comm{J_i^x}{J_j^z}=0$ 
and $\comm{J_i^y}{J_j^z}=0$ when $i \neq j$. In accordance with previous results 
we also keep only the $zz$ term among the three exchange interaction terms. 
Additionally, we neglect the hyperfine interaction, as it was  found, at least within mean field approximation, not to cause a significant difference in the $B_x-T$ phase diagram in the vicinity of the classical phase transition \cite{Bitko_Aeppli_1996_PhysRevLett.77.940}. The result is the effective Hamiltonian, given in Eq. \eqref{eq:rearranged-hamiltonain-eff} in the main text, which acts on the full 17-dimensional Hilbert space of the electronic angular momentum. The 2-state Ising space is created anew at each MC operation by determining the two lowest energy states and choosing one of them by the MC rules.

\section{Numerical Methods}In order to construct approximate many-body states
of the full system, we diagonalize the single-site Hamiltonian for each of the
sites, given some arbitrary initial set of magnetic moments. Using the magnetic
moments obtained for each spin, the fields \eqref{eq:effective-B} and ergo single-site Hamiltonians \eqref{eq:single-site-hamiltonian} are updated, and the
process is repeated until convergence.
Convergence is assumed when the absolute difference between the assigned magnetic moment and the magnetic moment dictated by the local field, averaged over all sites, is smaller than $\epsilon_{\text{tol}}=5\times 10^{-3} \left[\mu_B\right]$.
Within this process, each spin is
assumed to be either "up" or "down", as set by the MC simulation, and only the magnitude of its magnetic moment is adjusted.  The process is performed
following each MC spin-flip.  In effect we neglect quantum many-body
effects such as entanglement, and instead consider each ion separately.
Nevertheless, the single ion is treated exactly by diagonalization of its
Hamiltonian in a manner that is self-consistent with all other ions.
This method is somewhat similar to the approach described in Ref.~\cite
{Piatek_nonequilibrium_2014} under the name inhomogeneous mean-field (iMF),
with one important difference. We do not replace the $J^z$ operators in the
Hamiltonian with their \protect \textit {thermal} averages but with their
quantum mechanical expectation values. Thus we allow the MC simulation to
sample thermal fluctuations which are required for the described mechanism to come into effect.

The determination of the single-site energy and magnetic moment is performed as follows:
The Hamiltonian \eqref{eq:single-site-hamiltonian} is diagonalized numerically and its two low energy levels are designated $\ket{\alpha}$ and $\ket{\beta}$ such that $E_{\alpha}<E_{\beta}$.
Next, the states $\ket{\alpha}$ and $\ket{\beta}$ are identified as $\ket{\uparrow}$ or $\ket{\downarrow}$ according to their $\expval{J^z}$ in the following manner: If $\expval{J^z}{\alpha}>\expval{J^z}{\beta}$ then $\ket{\alpha} \equiv \ket{\uparrow}$ and $\ket{\beta} \equiv \ket{\downarrow}$ and vise versa otherwise.
Hybridization between the $\up$ and $\down$ states is suppressed due to the hyperfine interaction \cite{Schechter_Stamp_2005_PhysRevLett_95_267208,Schechter_Stamp_2008_PhysRevB.78.054438}, which we approximately take into account by introducing an artificial longitudinal field in the determination of $\expval{J^z}$ that ensures the $\up$ and $\down$ are not significantly hybridized by the transverse field: For each applied local field, $(B_x,B_y,B_z)$, the $B_z$ component is replaced by $1.1 \times \sqrt{B_x^2 + B_y^2}$ for the purpose of obtaining the magnetic moment. The effect of this process can be seen in Supplementary Figure~\ref{fig:mag-moment-vs-b}.
In practice the energy and magnetic moment are calculated as described on a fixed grid of $B_x$, $B_y$ and $B_z$ from which they are linearly interpolated during the simulation.

\begin{figure}
	\centering
	\includegraphics[width=0.55\textwidth]{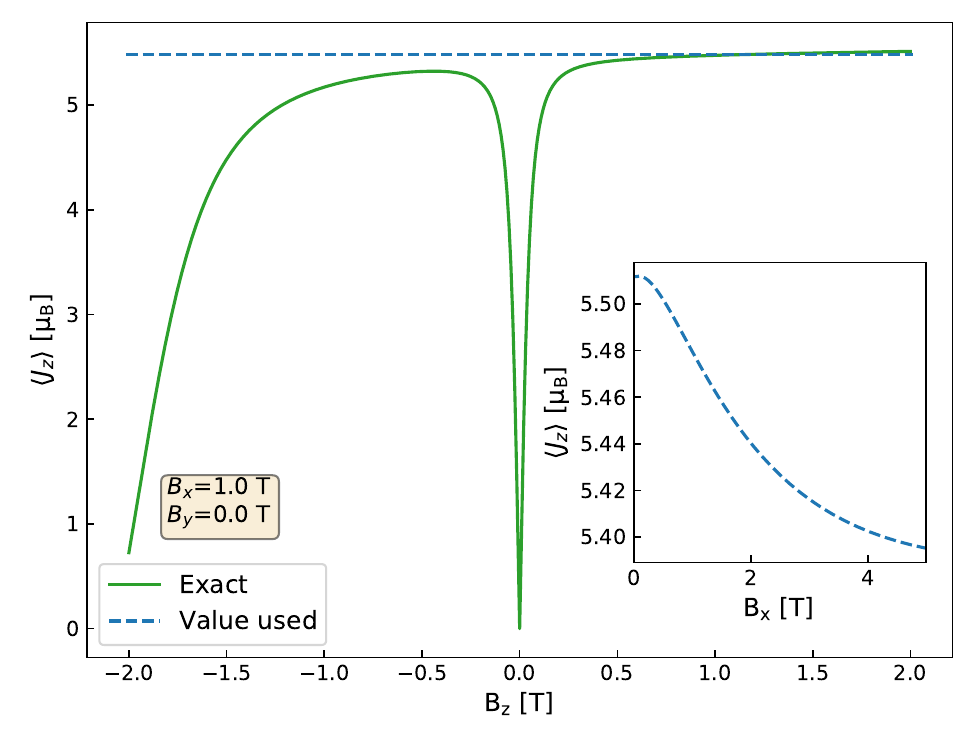}
	\caption{The magnetic moment $\expval{J^z}$ of the "up" state of a single \Ho{} ion as a function of the applied $B_z$ magnetic field, for an applied $B_x=\SI{1}{\tesla}$ field. The solid green line shows the "exact" result (without the hyperfine interaction), which shows significant hybridization when $B_z$ is smaller than $B_x=\SI{1}{\tesla}$. The dashed blue line shows the value used in this work, chosen as described in the text.
		The inset shows how this value changes as $B_x$ is varied.}
	\refstepcounter{SIfig}\label{fig:mag-moment-vs-b}
\end{figure}

Periodic boundary conditions are used, and the dipolar interaction $V_{ij}^{\mu\nu}$ is calculated using the Ewald summation method without a demagnetization term \cite{Tam_spin-glass_2009,Wang_Holm_2001_J._Chem._Phys._1156351}.

We use the parallel tempering Monte Carlo method
\cite{Hukushima_1996_JPSJ.65.1604}. To determine
the transitions at a given $x$ and $B_x$ we use the finite-size correlation
length \cite{Ballesteros_2000_PhysRevB.62.14237,Tam_spin-glass_2009},
\begin{equation}\label{eq:correlation-length} \xi_L = \frac{1}{2
		\sin(k_{\text{min}/2})} \left[
	\frac{\left[\expval{m^2(0)}_T\right]_{\text{av}}}{\left[\expval{m^2(\boldsymbol{k}_{\text{min}})}_T\right]_{\text{av}}}
	- 1 \right]^{\frac{1}{2}} \end{equation} where \begin{equation}\label{eq:mk}
	m(\boldsymbol{k})=\frac{1}{N} \sum_{i=1}^N \expval{J_i^z}_{\psi} \exp(-
	i\boldsymbol{k}\cdot \boldsymbol{R}_i).  \end{equation} Here $\expval{.}_T$
refers to a thermal average, $\left[ . \right]_{\text{av}}$ to a disorder
average and $\expval{.}_{\psi}$ to a quantum mechanical expectation value.
$\boldsymbol{R}_i$ is the location of the site $i$ and
$\boldsymbol{k}_{\text{min}}=(\frac{2\pi}{L},0,0)$. The finite-size correlation
length divided by the linear system size $L$ has a known scaling form,
\begin{equation}\label{eq:FSS-correlation-length-L} \frac{\xi_L}{L} \sim
	\tilde{X}\left(L^{1/\nu}(T-T_c)\right) \end{equation} so that for $T=T_c$ it is
independent of the system size $L$. Hence, if a transition exists, curves for
different system sizes should cross at the critical temperature. We simulate
systems of linear sizes $L=6,7,8$. To find $T_c$ we assume the scaling function
\eqref{eq:FSS-correlation-length-L} can be approximated by a third-order
polynomial close to the critical point: $f(x)=p_0 + p_1 x + p_2 x^2 + p_3 x^3$
(where $x=L^{1/\nu}(T-T_c)$), and perform a global fit for the six free
parameters, $p_0 \ldots p_3$, $\nu$ and $T_c$ using the Levenberg-Marquardt
algorithm. All 17 individual plots showing the analysis are presented in Supplementary Figures~\ref{fig:fss-odd-included}-\ref{fig:fss-odd-excluded-dilute}.
Statistical errors are estimated using the bootstrap method
\cite{Katzgraber_Young_2006_PhysRevB.73.224432}.
See Supplementary Tables \ref{table:sim-params-x-1}-\ref{table:sim-params-x-10} for numerical results with statistical errors.

Equilibration is verified by logarithmic binning of the data, i.e. the
simulation time in terms of MC sweeps is successively increased by a
factor of 2, and observables are averaged over that time. When all observables
of interest in three consecutive bins agree within error bars, the simulation is
deemed equilibrated \cite{Katzgraber_Young_2006_PhysRevB.73.224432}.

For each value of $x$ or $B_x$ the simulation is performed twice: Once with the
off-diagonal dipolar terms included in the Hamiltonian
\eqref{eq:rearranged-hamiltonain-eff} and once with these terms omitted so that internal transverse fields are artificially suppressed, i.e. $B^x_i=B_x$ and $B^y_i=0$ for all $i$ in \eqref{eq:effective-B}.
The self-consistent calculation for the magnetic moments is performed in both cases, as a means to establish its validity.

\section{Simulation Parameters}
Supplementary tables~\ref{table:sim-params-x-1}-\ref{table:sim-params-x-10} present the parameters and the results of the Monte Carlo simulations performed for this work.

\begin{table}[H]
	\caption{Simulation parameters at $x=1$ for different transverse fields $B_x$ and system sizes $L$, with ODD terms included and excluded. The equilibration/measurement times are $2^b$ Monte Carlo sweeps. $T_\text{min}$ [$T_\text{max}$] is the lowest [highest] temperature used and $N_T$ is the number of	temperatures. $N_{sa}$ is the number of independent runs.
		All simulations use $\Jex=\SI{1.16}{\milli\kelvin}$.}
	\label{table:sim-params-x-1}
	\centering
	\begin{tabular}{ c	c c c c c c c c c } 
		\hline\hline
		ODD terms & $B_x\ [\rm{T}]$ & L & b & $T_{\text{min}}\ [\rm{K}]$ & $T_{\text{max}}\ [\rm{K}]$ & $N_T$ & $N_{sa}$ & $T_c\ [\rm{K}]$ & $\nu$ \\ [0.5ex] 
		\hline
		Included & 0.0 & 6,7,8 & 10 & 1.528 & 1.628 & 24 & 50 & 1.5735(4) & 0.65(1) \\ 
		\hline
		Excluded & 0.0 & 6,7 & 10 & 1.738 & 1.838 & 24 & 50 & \multirow{2}{*}{1.7868(3)} & \multirow{2}{*}{0.59(1)} \\ 
		Excluded & 0.0 & 8 & 11 & 1.738 & 1.838 & 24 & 50 &  &  \\ 
		\hline
		Included & 0.3 & 6,7,8 & 10 & 1.512 & 1.612 & 24 & 50 & 1.5668(4) & 0.61(1) \\ 
		Excluded & 0.3 & 6,7,8 & 10 & 1.739 & 1.839 & 24 & 50 & 1.7800(4) & 0.60(1) \\ 
		Included & 0.6 & 6,7,8 & 10 & 1.498 & 1.598 & 24 & 30 & 1.5529(5) & 0.61(1) \\ 
		Excluded & 0.6 & 6,7,8 & 10 & 1.727 & 1.827 & 24 & 30 & 1.7666(5) & 0.60(2) \\ 
		Included & 1.0 & 6,7,8 & 10 & 1.47 & 1.57 & 24 & 30 & 1.5226(5) & 0.64(2) \\ 
		Excluded & 1.0 & 6,7,8 & 10 & 1.705 & 1.804 & 24 & 30 & 1.7502(5) & 0.61(2) \\ 
		\hline
		Included & 1.5 & 6,7 & 10 & 1.42 & 1.52 & 24 & 30 & \multirow{2}{*}{1.4713(4)} & \multirow{2}{*}{0.56(1)} \\ 
		Included & 1.5 & 8 & 11 & 1.42 & 1.52 & 24 & 30 &  & \\
		\hline
		Excluded & 1.5 & 6,7 & 10 & 1.67 & 1.77 & 24 & 30 & \multirow{2}{*}{1.7275(5)} & \multirow{2}{*}{0.62(2)} \\ 
		Excluded & 1.5 & 8 & 11 & 1.67 & 1.77 & 24 & 30 &  &  \\
		\hline\hline
	\end{tabular}
\end{table}

\begin{table}[H]
	\caption{Simulation parameters at $B_x=0$ for different dilutions $x\leq 1$ and system sizes $L$, with ODD terms included and excluded. The equilibration/measurement times are $2^b$ Monte Carlo sweeps. $T_\text{min}$ [$T_\text{max}$] is the lowest [highest] temperature used and $N_T$ is the number of	temperatures. $N_{sa}$ is the number of independent runs.}
	\label{table:sim-params-x-10}
	\centering
	\begin{tabular}{ c c c c c c c c c c c c } 
		\hline\hline
		$x$ & ODD terms & $\Jex\ [\rm{mK}]$ & L & b & $T_{\text{min}}\ [\rm{K}]$ & $T_{\text{max}}\ [\rm{K}]$ & $N_T$ & $N_{sa}$ & $T_c\ [\rm{K}]$ & $\nu$ \\ [0.5ex] 
		\hline
		1.0 & Included & 1.16 & 6,7,8 & 10 & 1.528 & 1.628 & 24 & 50 & 1.5735(4) & 0.65(1) \\ 
		0.83 & Included & 1.16 & 6,7,8 & 10 & 1.04 & 1.54 & 24 & 200 & 1.302(2) & 0.67(4) \\ 
		0.67 & Included & 1.16 & 6,7,8 & 10 & 0.78 & 1.28 & 24 & 350 & 1.044(2) & 0.66(4) \\ 
		\hline
		0.46 & Included & 1.16 & 6 & 11 & 0.39 & 0.89 & 24 & 1000 & \multirow{2}{*}{0.656(2)} & \multirow{2}{*}{0.70(3)} \\
		0.46 & Included & 1.16 & 7,8 & 12 & 0.39 & 0.89 & 24 & 1000 &  &  \\ [0.5ex]
		\hline
		1.0 & Excluded & 3.91 & 6,7,8 & 10 & 1.463 & 1.563 & 24 & 30 & 1.5177(5) & 0.59(1) \\ 
		0.83 & Excluded & 3.91 & 6,7,8 & 10 & 0.991 & 1.491 & 24 & 200 & 1.241(2) & 0.68(4) \\ 
		\hline
		0.67 & Excluded & 3.91 & 6,7 & 10 & 0.724 & 1.224 & 24 & 350 & \multirow{2}{*}{0.978(2)} & \multirow{2}{*}{0.64(4)} \\ 
		0.67 & Excluded & 3.91 & 8 & 11 & 0.724 & 1.224 & 24 & 350 &  &  \\ 
		\hline
		0.46 & Excluded & 3.91 & 6 & 11 & 0.354 & 0.854 & 24 & 1000 & \multirow{3}{*}{0.594(2)} & \multirow{3}{*}{0.65(2)} \\ 
		0.46 & Excluded & 3.91 & 7 & 12 & 0.354 & 0.854 & 24 & 1000 &  &  \\ 
		0.46 & Excluded & 3.91 & 8 & 13 & 0.354 & 0.854 & 24 & 1000 &  &  \\ 
		\hline\hline
	\end{tabular}
\end{table}

\section{Finite-Size Scaling analysis}
\begin{figure}[H]
	\centering
	\includegraphics[width=0.45\textwidth]{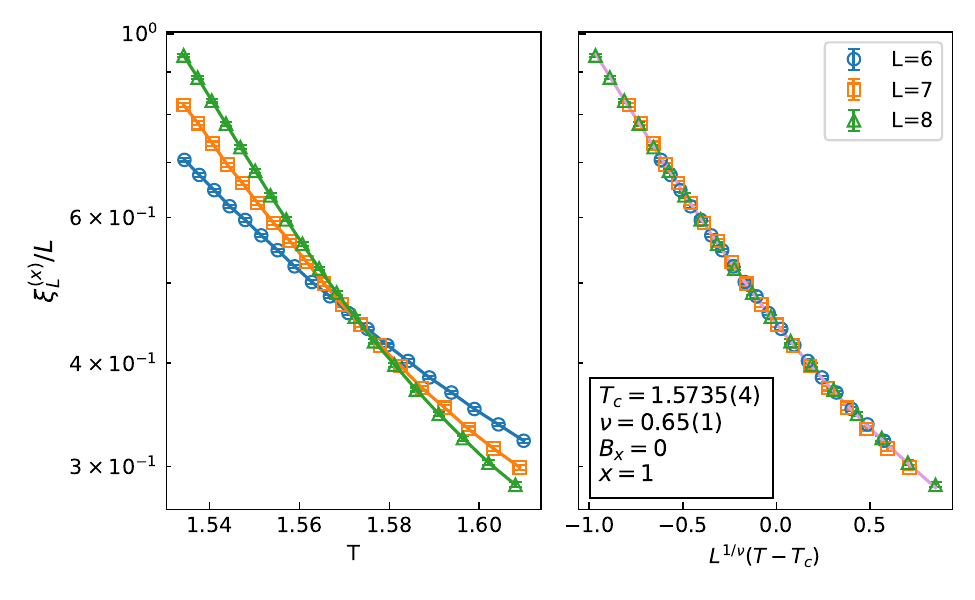}
	\includegraphics[width=0.45\textwidth]{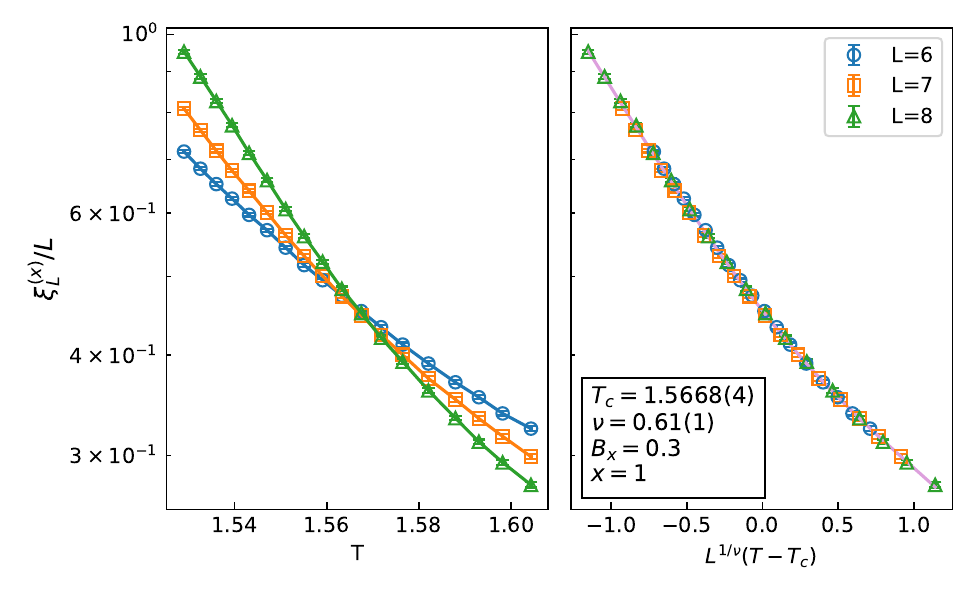}\\
	\includegraphics[width=0.45\textwidth]{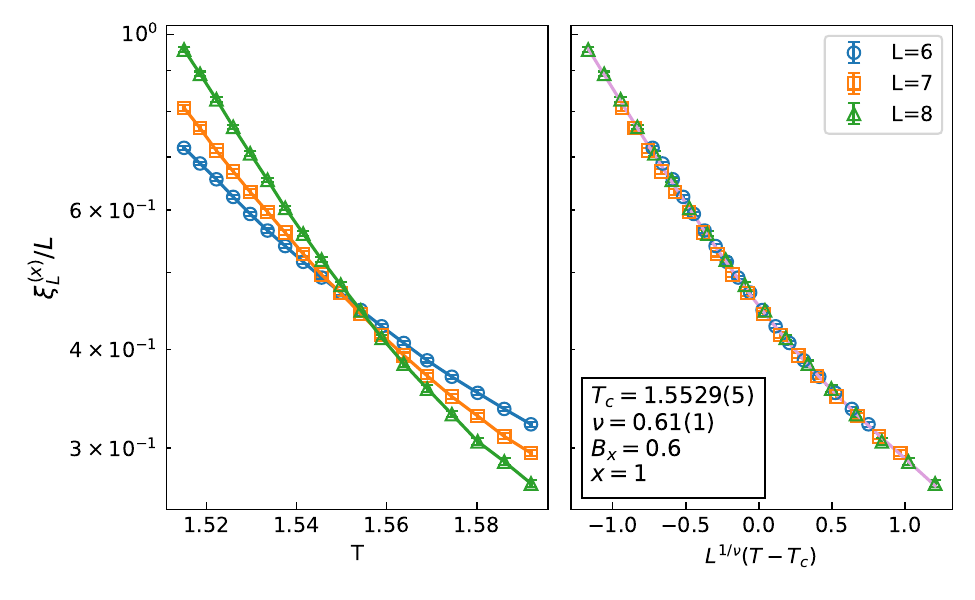}
	\includegraphics[width=0.45\textwidth]{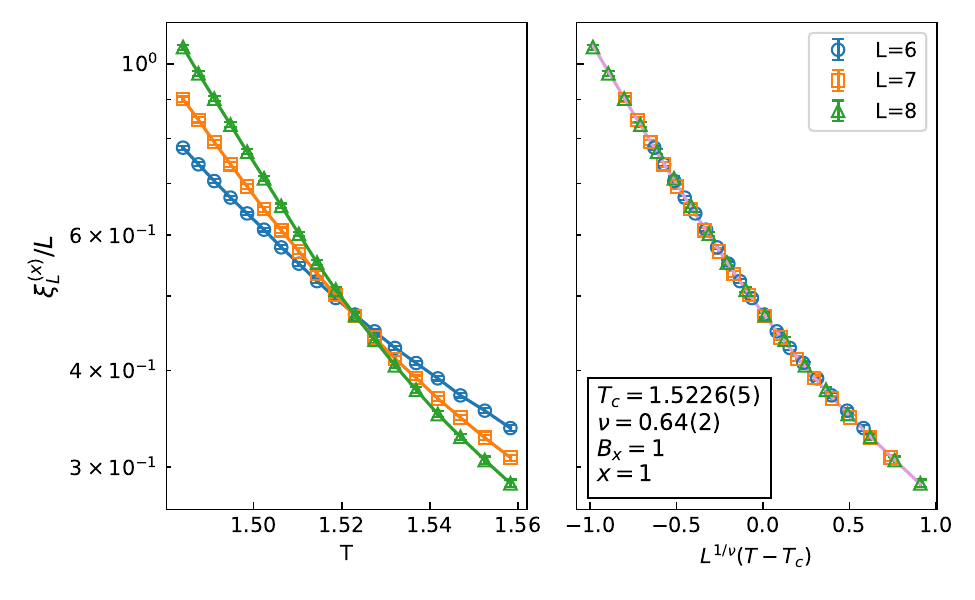}\\
	\includegraphics[width=0.45\textwidth]{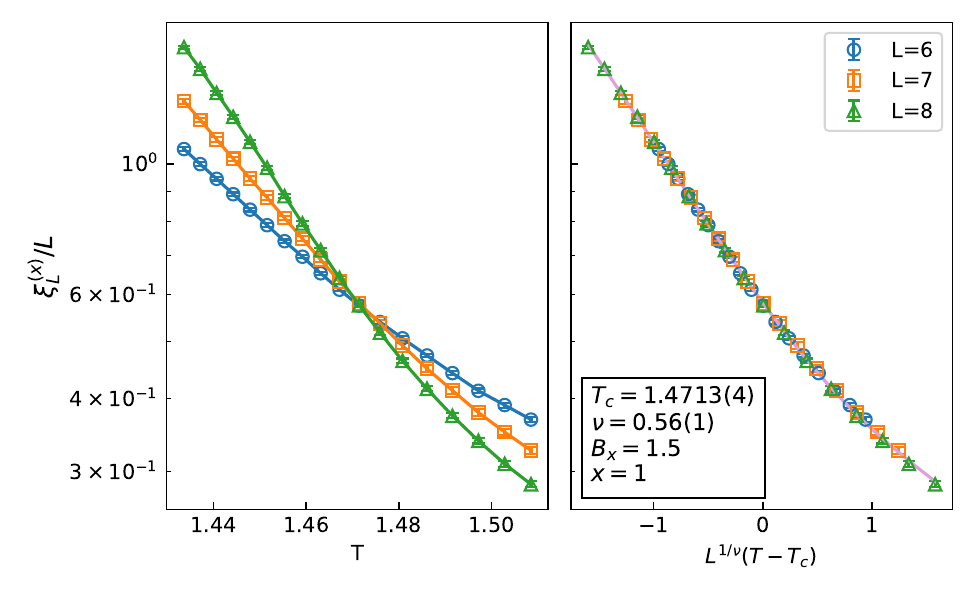}
	\caption{Finite size scaling results for simulations with ODD terms included at $x=1$ and $\Jex=\SI{1.16}{\milli\kelvin}$.
		Each graph is divided in two: the left side shows the raw results with the curves for different system sizes crossing at $T_c$ and the right side shows the same results with the $T$ axis rescaled showing a collapse onto a universal curve.
		The parameters for each simulation are listed within the figures.}
	\refstepcounter{SIfig}\label{fig:fss-odd-included}
\end{figure}

\begin{figure}[H]
	\centering
	\includegraphics[width=0.45\textwidth]{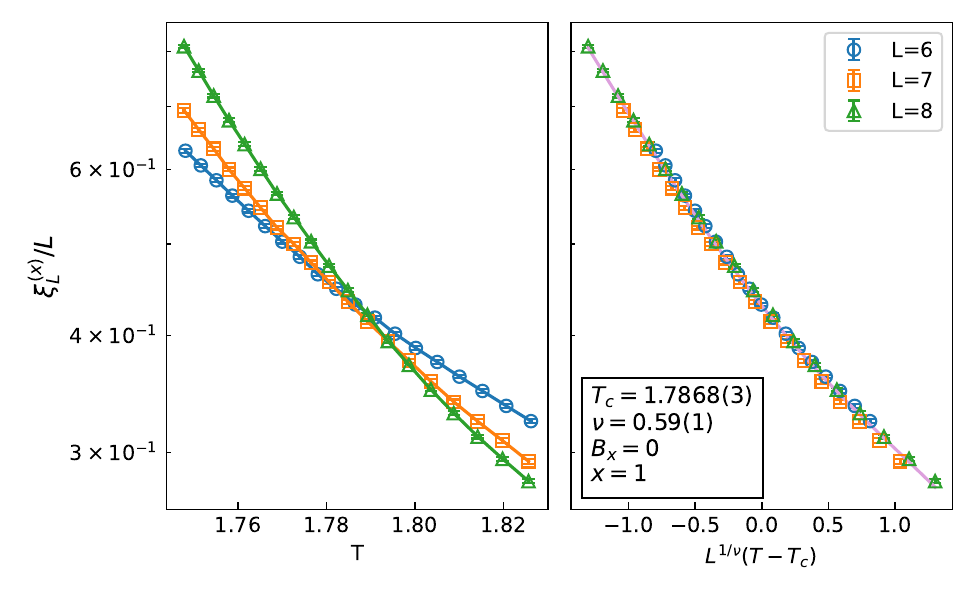}
	\includegraphics[width=0.45\textwidth]{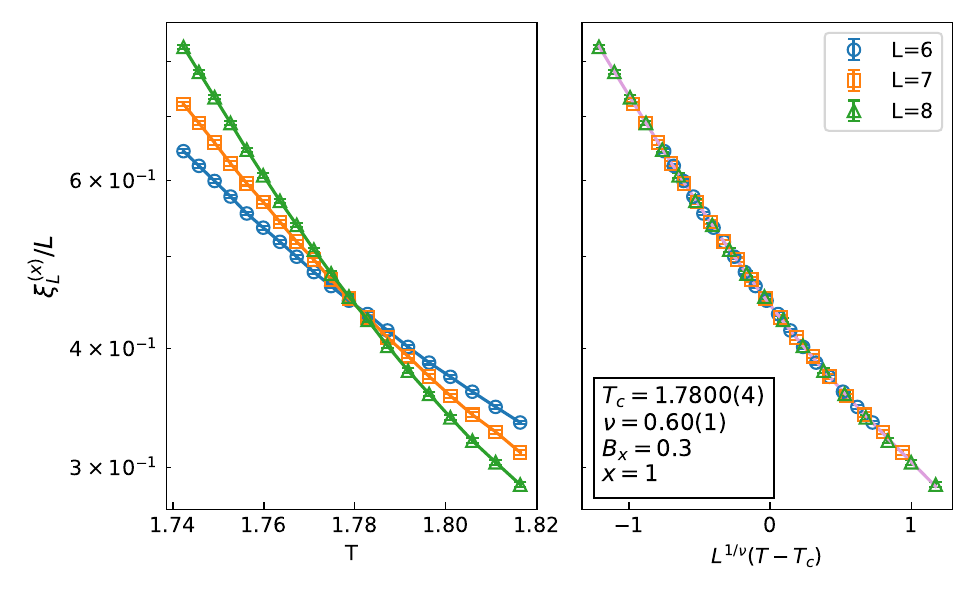}\\
	\includegraphics[width=0.45\textwidth]{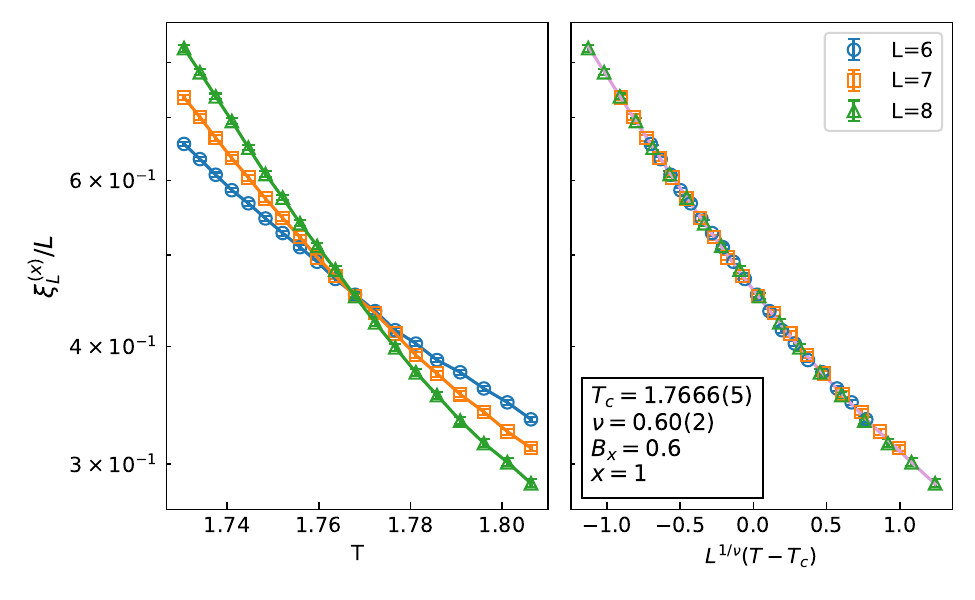}
	\includegraphics[width=0.45\textwidth]{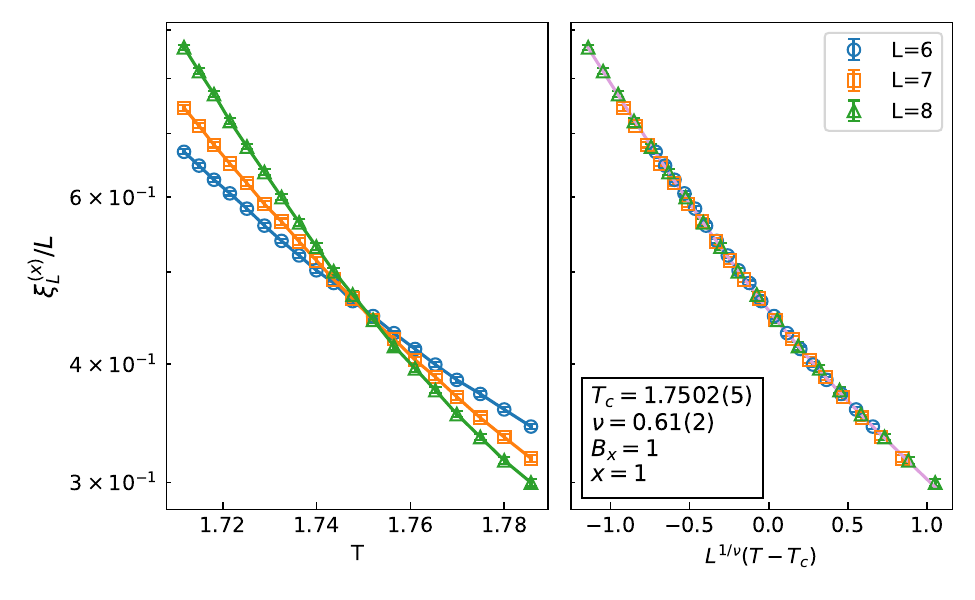}\\
	\includegraphics[width=0.45\textwidth]{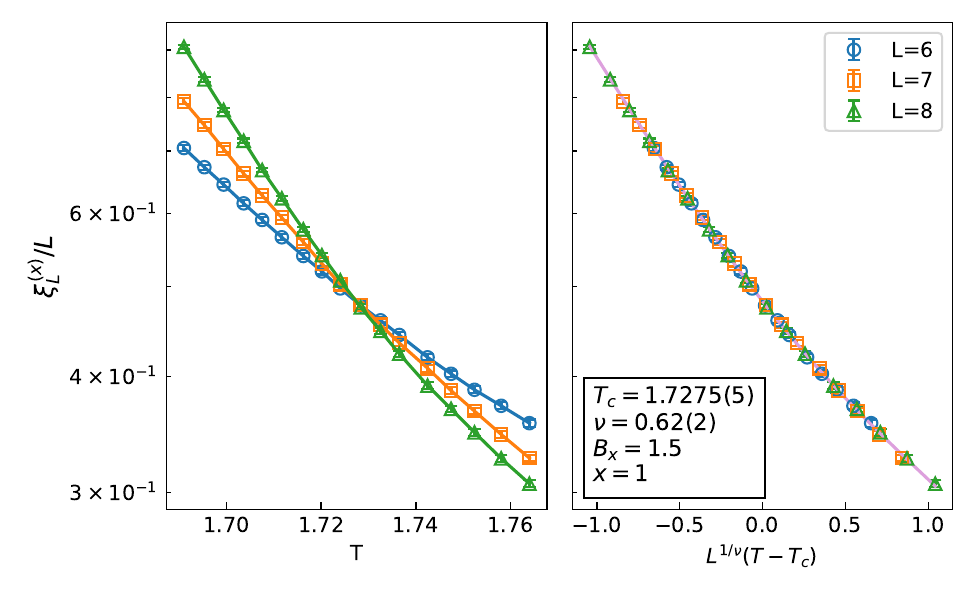}
	\caption{Finite size scaling results for simulations with ODD terms excluded.
		Each graph is divided in two: the left side shows the raw results with the curves for different system sizes crossing at $T_c$ and the right side shows the same results with the $T$ axis rescaled showing a collapse onto a universal curve.
		The parameters for each simulation are listed within the figures.}
	\refstepcounter{SIfig}\label{fig:fss-odd-excluded}
\end{figure}

\begin{figure}[H]
	\centering
	\includegraphics[width=0.45\textwidth]{figures/fit_0.0_false_6_7_8_low_jex_dilution_1.0_.pdf}
	\includegraphics[width=0.45\textwidth]{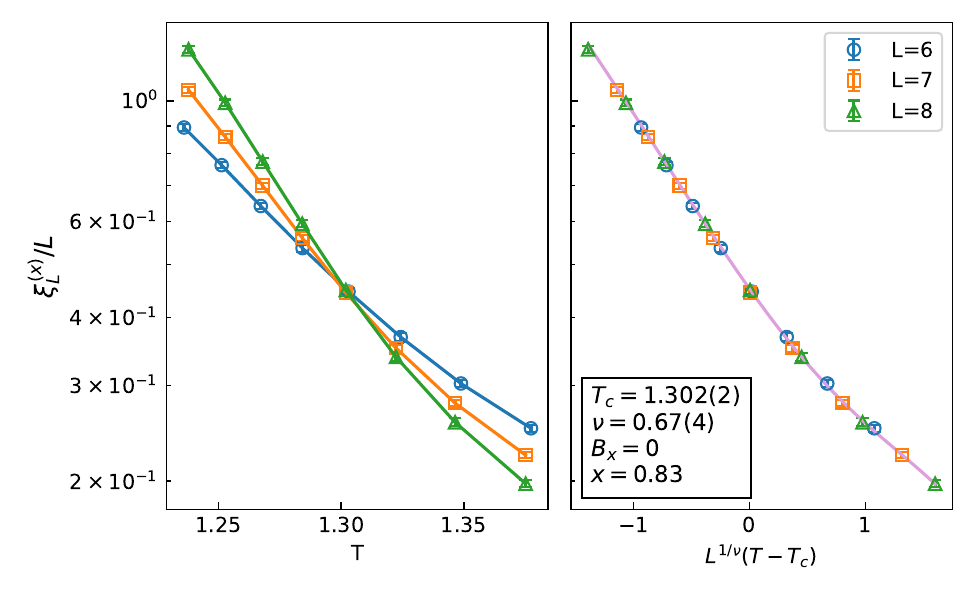}\\
	\includegraphics[width=0.45\textwidth]{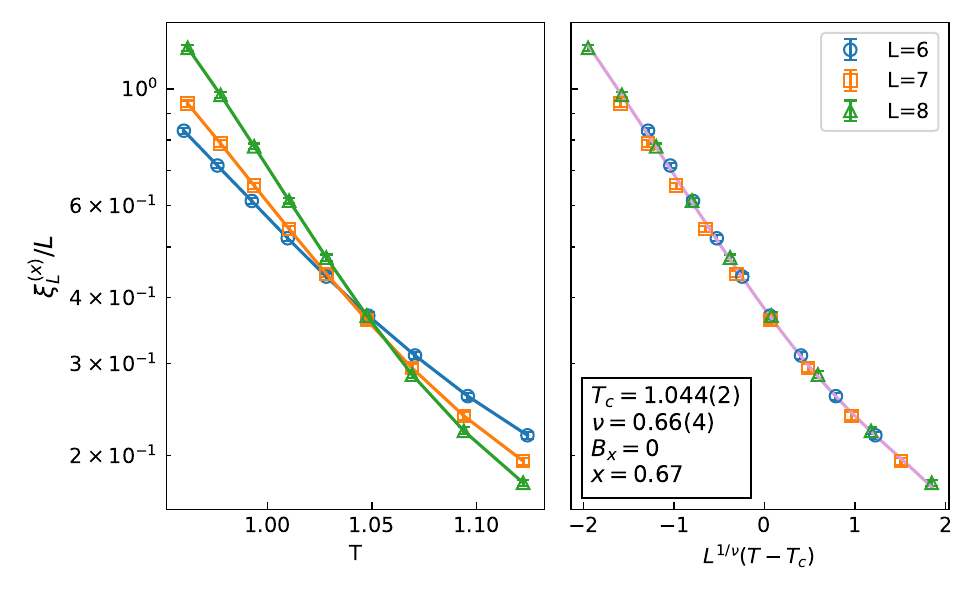}
	\includegraphics[width=0.45\textwidth]{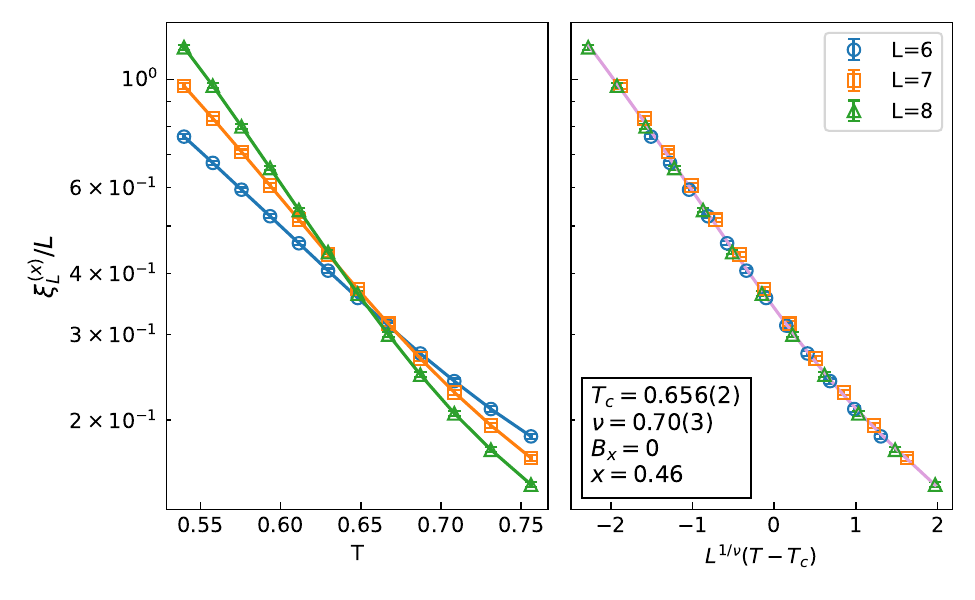}
	\caption{Finite size scaling results for simulations with ODD terms included at $x \leq 1$ and $\Jex=\SI{1.16}{\milli\kelvin}$.
		Each graph is divided in two: the left side shows the raw results with the curves for different system sizes crossing at $T_c$ and the right side shows the same results with the $T$ axis rescaled showing a collapse onto a universal curve.
		The parameters for each simulation are listed within the figures.}
	\refstepcounter{SIfig}\label{fig:fss-odd-included-dilute}
\end{figure}

\begin{figure}[H]
	\centering
	\includegraphics[width=0.45\textwidth]{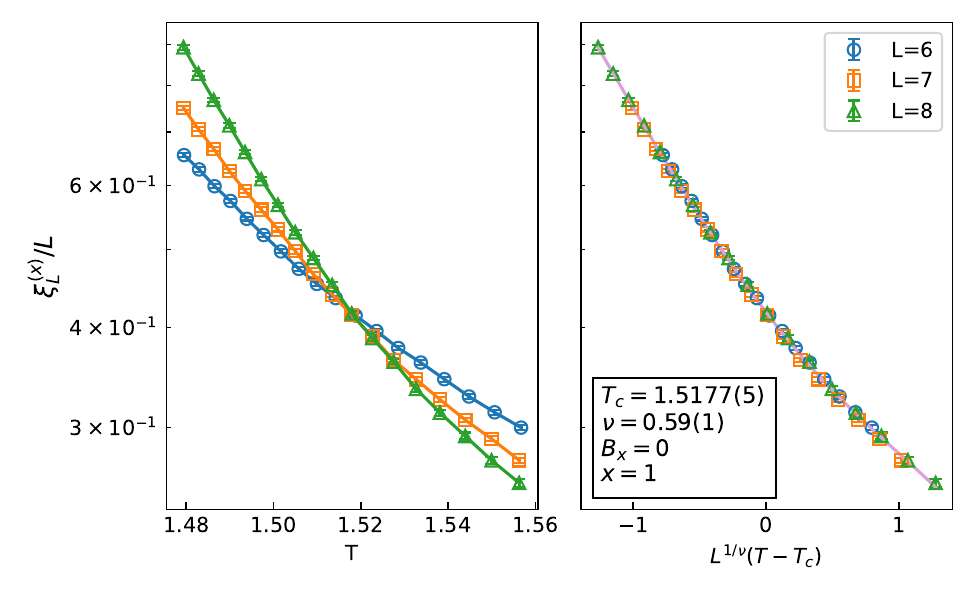}
	\includegraphics[width=0.45\textwidth]{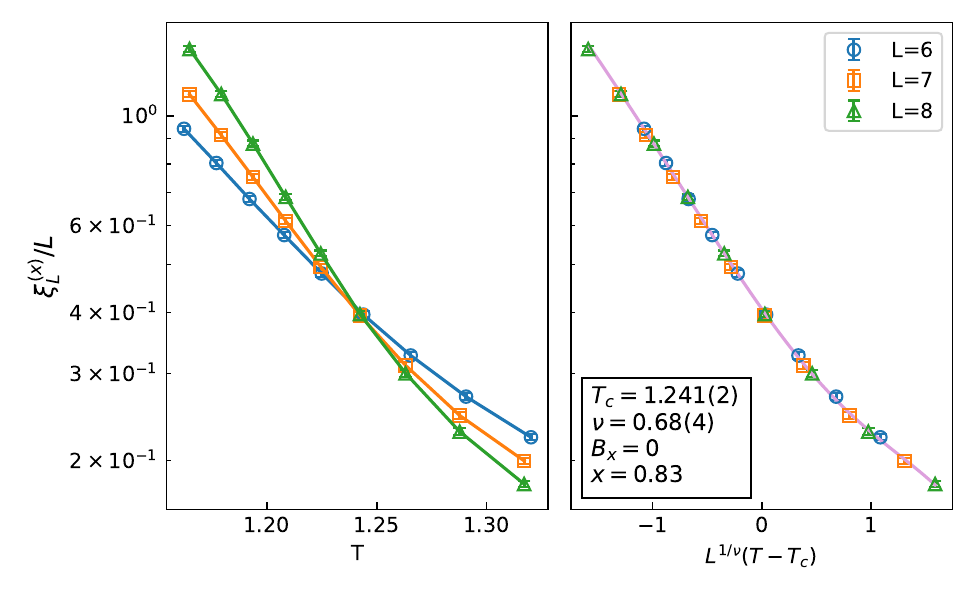}\\
	\includegraphics[width=0.45\textwidth]{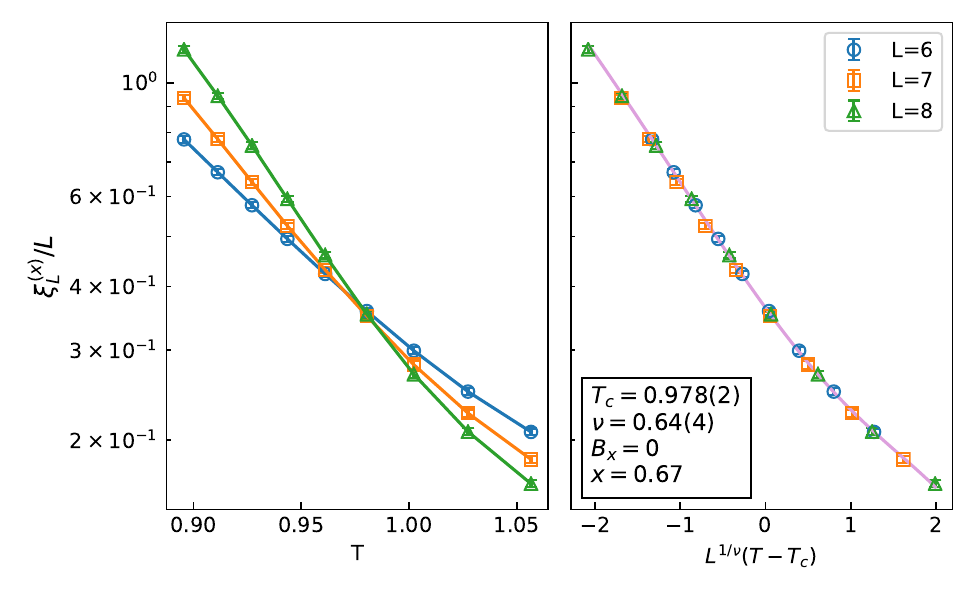}
	\includegraphics[width=0.45\textwidth]{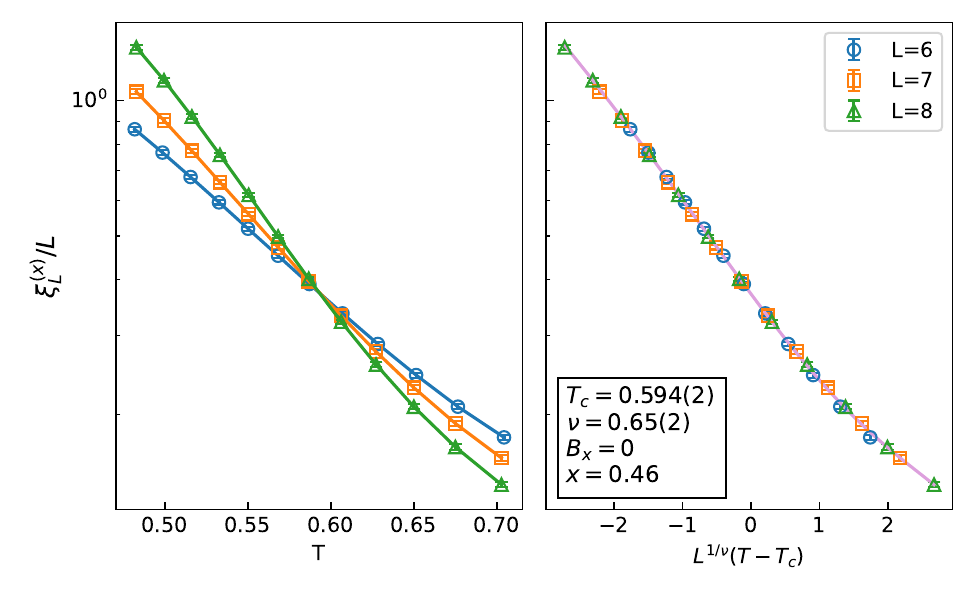}
	\caption{Finite size scaling results for simulations with ODD terms excluded at $x \leq 1$ and $\Jex=\SI{3.91}{\milli\kelvin}$.
		Each graph is divided in two: the left side shows the raw results with the curves for different system sizes crossing at $T_c$ and the right side shows the same results with the $T$ axis rescaled showing a collapse onto a universal curve.
		The parameters for each simulation are listed within the figures.}
	\refstepcounter{SIfig}\label{fig:fss-odd-excluded-dilute}
\end{figure}

This section presents the results of the finite size scaling analysis used to obtain $T_c$ from each simulation.
For each simulation, the value of $B_x$ and $x$ is set, and then a range of temperatures around $T_c$ is simulated for several system sizes. These simulations are used to obtain the averages required to calculate the finite-size correlation length $\xi_L$ which is used in the finite-size scaling method to obtain an exact estimate of $T_c$. Supplementary Figures~\ref{fig:fss-odd-included}-\ref{fig:fss-odd-excluded-dilute} show the finite-size scaling results.
Despite expected corrections to scaling arising from the relatively small system sizes, all figures show reasonable collapse to a universal curve.

\section{Estimation of the Effective Interaction}
The excess change in the energy of the system resulting from the offdiagonal dipolar (ODD) interactions can be viewed, to some extent, as a change in the effective longitudinal (zz) pair interactions. This effective change in the pair interactions is specific to each pair, is dependent on the specific spatial configuration at finite concentration, on transverse magnetic field, and to some extent on the specific configuration of all spins in the system. Yet, it is useful to calculate it in some specific configurations to allow estimation of the contribution of the offdiagonal terms to the energy of the system, and thus to $T_c$.
To calculate the effective interaction due to the ODD mechanism, we take as one example a ferromagnetic system of all spins, except two, in the \textit{up} state. We then calculate the energy of the \textit{full} system for four different configurations of these 2 spins, $E(\uparrow\uparrow)$, $E(\uparrow\downarrow)$, $E(\downarrow\uparrow)$, $E(\downarrow\downarrow)$, where the energy is calculated using $H_{\text{eff}}$ given in Eq. \eqref{eq:rearranged-hamiltonain-eff} in the main text, with magnetic moments calculated self consistently as described in the text.
The interaction between the two spins is then given by
\begin{equation*}
	J=\frac{1}{4}\left[E(\uparrow\uparrow) + E(\downarrow\downarrow) - E(\uparrow\downarrow) - E(\downarrow\uparrow) \right]
\end{equation*}
which is a result of the direct longitudinal dipolar interaction, the (nearest-neighbor) exchange interaction and the effective excess interaction due to the ODD mechanism.
By repeating the above calculation with and without offdiagonal dipolar terms and subtracting one from the other, we get an estimate of the excess interaction due to the ODD mechanism $J_{\text{eff}} \equiv J(\text{ODD included}) - J(\text{ODD excluded})$.

This effective interaction is calculated for pairs of spins along the x axis, along the z axis and for nearest neighbors (which can be seen in Fig. 2a in the main text).
Results are obtained for a system of linear size $L=14$, where dipolar interactions are calculated using the Ewald method.
For nearest neighbors the effective interaction is $J_{\text{eff}}=-31\ \mathrm{mK}$ (ferromagnetic),  for the two closest spins along the x axis it is $J_{\text{eff}}=14\ \mathrm{mK}$ (antiferromagnetic), and for the two closest spins along the z axis it is $J_{\text{eff}}=11\ \mathrm{mK}$ (antiferromagnetic).
These values amount to roughly 7-23\% of the standard longitudinal dipolar interactions of the respective pairs.
Similar results are obtained for an average over a random distribution of the spins.

\end{document}